\documentclass[sigconf]{acmart}

\usepackage{booktabs} 

\usepackage[ruled]{algorithm2e} 
\usepackage{natbib}
\setcitestyle{square, comma, numbers,sort&compress, super}
\usepackage{amsmath,amssymb,amsfonts}
\usepackage{balance}
\usepackage{algorithmic}
\usepackage{multirow}
\usepackage{textcomp}

\SetAlFnt{\small}
\SetAlCapFnt{\small}
\SetAlCapNameFnt{\small}
\SetAlCapHSkip{0pt}
\IncMargin{-\parindent}

\acmJournal{TWEB}
\acmVolume{9}
\acmNumber{4}
\acmArticle{39}
\acmYear{2010}
\acmMonth{3}
\copyrightyear{2009}

\setcopyright{acmlicensed}

\acmDOI{0000001.0000001}

\setcopyright{none}
\makeatletter
\def\@copyrightspace{\relax}
\makeatother
\renewcommand\footnotetextcopyrightpermission[1]{} 
\begin{document}
\title[Heartbeat Classificatoin using Time-Frequency Distribution]{Heartbeat Classification in Wearables Using Multi-layer Perceptron and Time-Frequency Joint Distribution of ECG}

\author{Anup Das}
\orcid{1234-5678-9012-3456}
\affiliation{%
  \institution{Department of ECE, Drexel University}
  \streetaddress{3141 Chestnut Street}
  \city{Philadelphia}
  \state{PA}
  \postcode{19104}
  \country{USA}}
\email{anup.das@drexel.edu}

\author{Francky Catthoor}
\authornote{Dr. Francky Catthoor is also associated with the ESAT department of KU Leuven, Belgium.}
\affiliation{%
  \institution{IMEC}
  \city{Leuven}
  \country{Belgium}
}

\author{Siebren Schaafsma}
\affiliation{%
 \institution{Stichting Imec Nederland}
 \streetaddress{Kapeldreef 75, 3001}
 \city{Eindhoven}
 \country{Netherlands}}

\begin{abstract}
	Heartbeat classification using electrocardiogram (ECG) data is a vital assistive technology for wearable health solutions. We propose heartbeat feature classification based on a novel sparse representation using time-frequency joint distribution of ECG. Fundamental to this is a multi-layer perceptron, which incorporates these signatures to detect cardiac arrhythmia. This approach is validated with ECG data from MIT-BIH arrhythmia database. Results show that our approach has an average 95.7\% accuracy, an improvement of 22\% over state-of-the-art approaches. Additionally, ECG sparse distributed representations generates only 3.7\% false negatives, reduction of 89\% with respect to existing ECG signal classification techniques.
\end{abstract}

%

\copyrightyear{2018} 
\acmYear{2018} 
\setcopyright{acmcopyright}
\acmConference[CHASE '18]{ACM/IEEE International Conference on Connected Health: Applications, Systems and Engineering Technologies}{September 26--28, 2018}{Washington, DC, USA}
\acmBooktitle{ACM/IEEE International Conference on Connected Health: Applications, Systems and Engineering Technologies (CHASE '18), September 26--28, 2018, Washington, DC, USA}
\acmPrice{15.00}
\acmDOI{10.1145/3278576.3278598}
\acmISBN{978-1-4503-5958-0/18/09}


%
%


\maketitle


\section{Introduction}

\label{sec:introduction}
With the growing maturity of semiconductor process technology, miniature sensors are increasingly being integrated into wearable devices, facilitating vital health monitoring of humans. One important sensor in this context is the electrocardiogram (ECG) sensor, recording electrical activities associated with the human heart. ECG signals are useful to detect (1) cardiac arrhythmia, related to abnormal heart rhythms; (2) ischemia, related to poor blood flow to the heart muscles, (3) heart abnormalities such as enlarged heart; and (4) past heart attacks. In this paper, we focus on accurate classification of heartbeats leading to accurate detection of cardiac arrhythmia, which is a key technology enabler for wearable health solutions. 

\begin{figure}[t]
	\centering
	\centerline{\includegraphics[width=0.4\columnwidth]{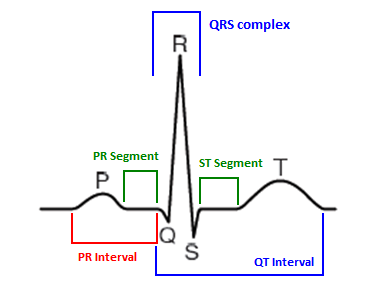}}
	\caption{QRS complex in ECG signal.}
	\label{fig:qrs_complex}
\end{figure}

Figure~\ref{fig:qrs_complex} shows the QRS complex in a ECG signal. This corresponds to the depolarization of the left and the right ventricles of a human heart. The \textbf{P, Q, R, S} and \textbf{T} waves occur in quick succession as shown in the figure. According to the publicly available ECG database released for cardiac arrhythmia (MIT-BIH arrhythmia database \cite{moody2001impact}), heartbeats can be classified into 23 different categories such as normal beats, atrial premature beats, superventicular escape beats, etc. based on the relative position and shape of these waves. However, according to the ANSI/AAMI/ISO EC57:1998/(R)2008 (American National Standard on testing and reporting performance results of cardiac rhythm and ST-segment measurement algorithms) \cite{association2008testing}, only five categories are sufficient -- (1) normal beats, (2) superventricular beats, (3) ventricular beats, (4) fusion beats and (5) unclassified beats. Additionally, arrhythmia detection is a direct function of these five heartbeat classes. The arrhythmia detection problem is therefore reduced to accurate classification of heartbeat classes. Although our approach considers all 23 classes, the classification accuracy is improved when a classifier is trained using only these 5 required classes.

Significant research have been conducted recently to autonomously classify ECG heartbeats (QRS peaks) \cite{de2004automatic,llamedo2011heartbeat,park2015arrhythmia,queiroz2015automatic,zhang2014heartbeat,al2016deep,li2016ecg,park2014pchd,martis2014computer,shufni2015ecg} (Section~\ref{sec:related_works} provides details of these approaches.). These studies have the following limitations that we address in this work. First, some of these studies are based on using designer specified deterministic QRS features, which in some cases can be incomplete. Machine learning provides a framework for automatic extraction of hidden features from ECG signals and is therefore gaining increasing popularity in this community. Second, existing machine learning based approaches use either time samples or frequency components for classification. We show that a time-frequency joint distribution captures time-varying frequency components, which is critical in anomaly detection of non-stationary and non-Gaussian signals such as speech and ECG. Third, most works fail to analyze the impact of noise on heartbeat classification accuracy, which is important, as noise from data acquisition systems are unavoidable.

In this paper we propose a two-phase approach to real-time heartbeat classification. Fundamental to this is an offline phase (phase I) that uses time-frequency joint distribution of QRS complexes and convert it into sparse distributed signatures that form the basis of classification using multi-layer perceptron. The trained classifier is then used online (phase II) to classify live ECG heartbeats detected using Pan \& Tompkin's Algorithm\cite{pan1985real}. Experiments conducted with the standard MIT-BIH arrhythmia database show that our approach is able to classify heartbeats with 80.7\% -- 100\% accuracy (average 95.7\%) for all heartbeat classes. Additionally, the proposed ECG sparse distributed signatures have high noise tolerance (by average 133\%) and low false negatives of only 3.7\%, an average 89\% lower compared to existing ECG representations.

\noindent{\textbf{\textbf{Contributions:}}} Following are our key contributions.
\begin{itemize}
	\item a two-phase approach for autonomous and real-time heartbeat classification;
	\item a time-frequency joint distribution based sparse distributed  representation (signature) for ECG signals;
	\item a multi-layer perceptron-based classifier to classify these signatures; and
	\item noise analysis of the proposed approach with Additive White Gaussian Noise.
\end{itemize}

The remainder of this paper is organized as follows. Related works are discussed in Section~\ref{sec:related_works}. The proposed two-phase approach is discussed in Section~\ref{sec:proposed_approach} and the problem formulation in Section~\ref{sec:problem_formulation}. Results are presented in Section~\ref{sec:results} and conclusions in Section~\ref{sec:conclusion}.

\section{Related works}
\label{sec:related_works}
Existing studies on heartbeat classification can be categorized into -- (1) classification using deterministic features and (2) classification using machine learning. Autonomous classification using deterministic heartbeat features have received research attention with the recent advances and wide-spread use of wearable devices such as the ECG band and smart watch. In \cite{de2004automatic}, heartbeats are classified using ECG morphology, heartbeat intervals, and RR-intervals, extracted  from ECG signals. This model is extended in \cite{llamedo2011heartbeat} with five additional features and in \cite{park2015arrhythmia} with ECG signal amplitude difference as a key feature. Although these techniques are based on features that are clinically understood, in some cases these can be incomplete.
Supervised machine learning in the form of classifiers are emerging as a popular alternative to the classical deterministic approaches due to the ability to autonomously identify and represent hidden features extracted from ECG signals. In \cite{zhang2014heartbeat}, support vector machine is used to classify ECG signals. A deep neural network based classifier is used as an alternative in \cite{al2016deep}. Other alternatives include the random forest approach of~\cite{li2016ecg} and the decision tree approach of~\cite{park2014pchd}. All these approaches use ECG time samples. There are also some studies using ECG frequency components. In \cite{martis2014computer}, coefficients of discrete wavelet transform are used for classification using a neural network. In \cite{shufni2015ecg}, ECG features are extracted using discrete Fourier transform of ECG time samples. As discussed in Section~\ref{sec:introduction}, none of these approaches consider time variation of the frequency components of ECG signals, which is essential to identify anomalies such as arrhythmia detection.

\section{Proposed Approach}
\label{sec:proposed_approach}
\begin{figure}[t]
	\centering
	\centerline{\includegraphics[width=0.9\columnwidth]{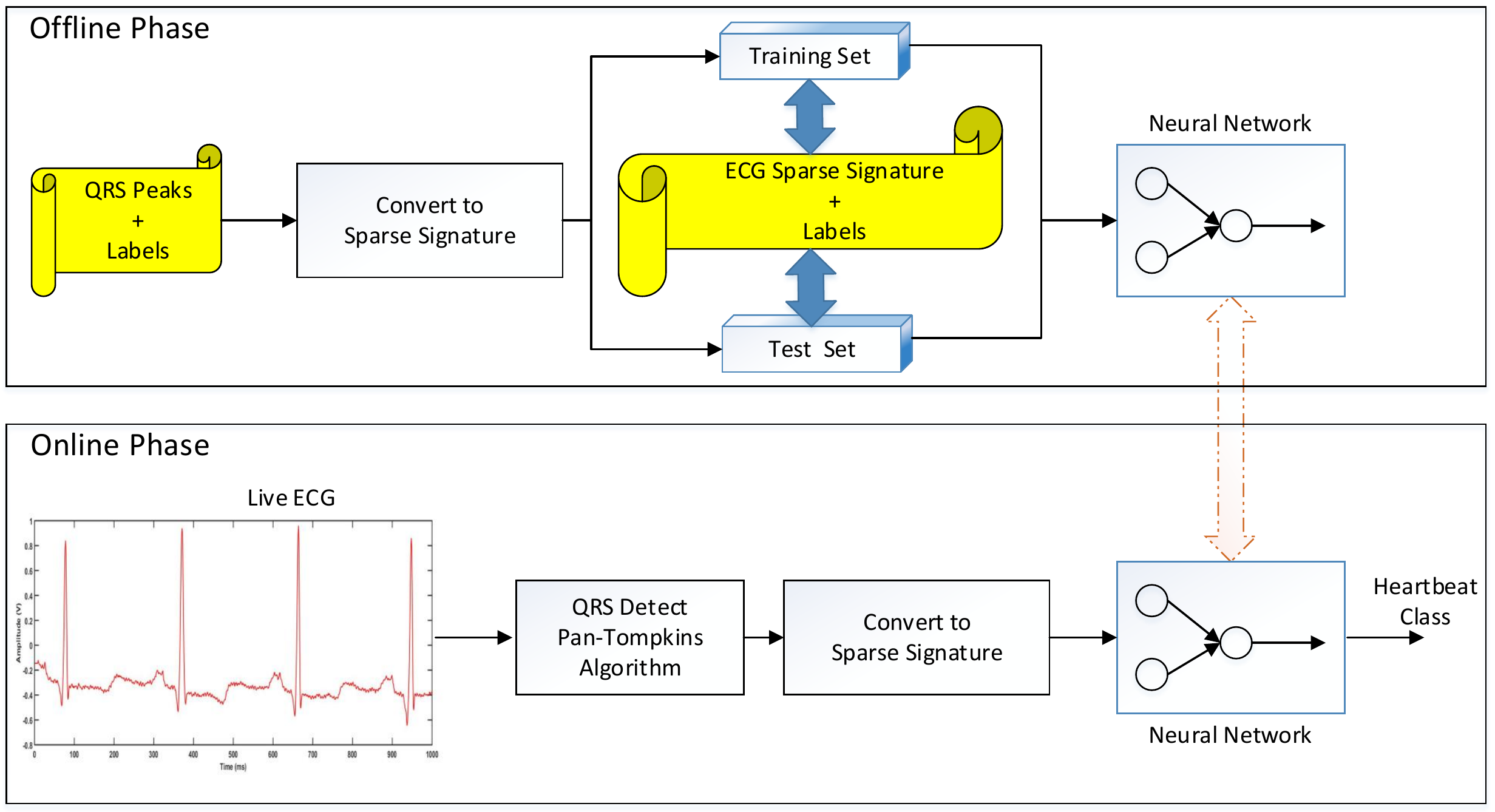}}
	\caption{Offline and online phases of the proposed approach.}
	\label{fig:proposed_approach}
\end{figure}

\begin{figure*}[t]
	\centering
	\centerline{\includegraphics[width=1.95\columnwidth]{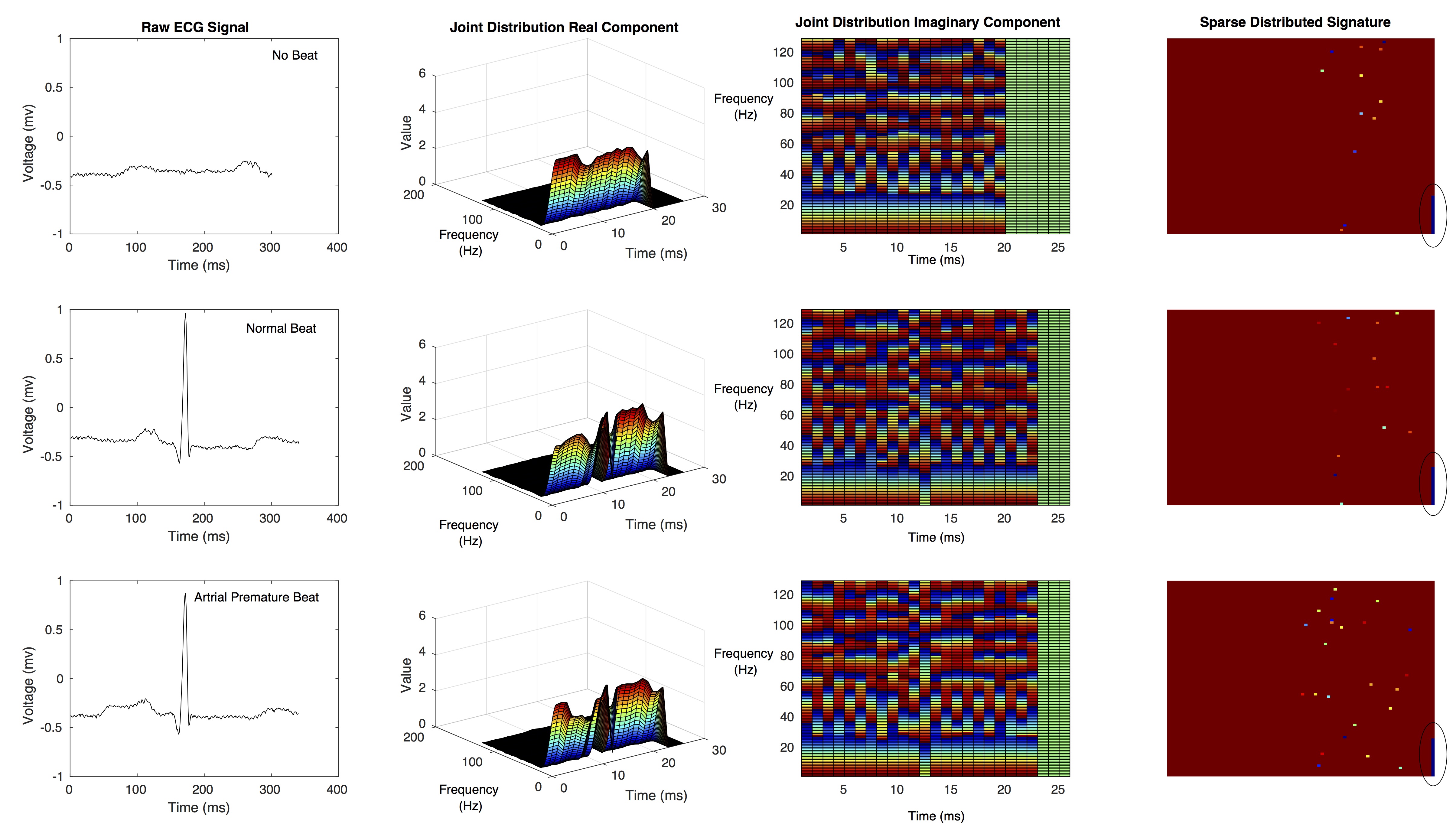}}
	\caption{ECG signal and plot for real and imaginary components of time-frequency joint distribution. Also shown are the sparse distributed signatures.}
	\label{fig:ecg_motvation}
\end{figure*}

Figure~\ref{fig:proposed_approach} outlines our proposed approach, comprising of an offline training phase followed by a online phase for real-time heartbeat classification. The offline training phase consists of training a deep neural network using labeled QRS peaks from the MIT-BIH arrhythmia database \cite{moody2001impact}. Each QRS peak is first transformed into its time-frequency representation. The ECG signal together with its real and imaginary components of time-frequency joint distribution are plotted in Figure~\ref{fig:ecg_motvation} for three scenarios -- (1) No Beat, (2) Normal Beat and (3) Atrial Premature Beat. As seen from Figure~\ref{fig:ecg_motvation}, the real and imaginary plots differ from each other for all three scenarios indicating the significance of both these components. We propose to combine these components to form unique signatures, which are shown on the rightmost subplots of Figure~\ref{fig:ecg_motvation}. These ECG signatures together with their original labels are then used for training the neural network as shown in Figure~\ref{fig:proposed_approach}.

The offline training phase is carried out on nvidia GPU using python machine learning library (theano \cite{bergstra2011theano}). The classifier used in this case is the multi-layer perceptron (MLP) with error back propagation. This choice is due to the fact that MLP are computationally less complex (important for real-time use), yet powerful tool and can be easily ported on a neuromorphic hardware \cite{indiveri2015neuromorphic}. Once trained, the neural network is validated using a test set. A design space exploration is then performed to determine dimensions of the neural network (hidden layers, number of neurons, etc). Online phase of the proposed approach uses ECG time samples to first detect QRS peaks using Pan \& Tompkins Algorithm \cite{pan1985real}. Detected peaks are then converted into ECG signatures and are classified using the trained neural network. This online phase can also be integrated within a ECG acquisition system \cite{konijnenburg201628,gradl2012real}.

\section{Problem Formulation}
\label{sec:problem_formulation}
\subsection{ECG Sparse Distributed Signatures }
The classical Fourier analysis assumes that signals are of infinite  duration or periodic in nature. However, non-stationary signals such as the ECG and speech are of short duration (transient), and change substantially over this duration. To accurately capture time varying frequency components of these transient signals, time-frequency joint distributions (such as short-time Fourier transform and wavelet transform) are frequently used. We use short-time Fourier transform of ECG signals due to its real-time properties. 

The short-time Fourier transform of a discrete-time signal $\small x(t)$ is given by the following equation \cite{almeida1994fractional}
\begin{equation}
\label{eq:eq01}
\footnotesize X(t,\omega) = \sum_{\tau=-\infty}^{\infty} h(t-\tau)\cdot x(\tau)\cdot e^{-j\omega\tau}
\end{equation}
where, $h(t)$ is the analysis window, which is narrow in time and frequency, and is normalized such that $\small h(0) = 1$. This equation is similar to the classical Fourier transform, except that $\small X(t,\omega)$ is now a function of both time and frequency, and represents only a local behavior of $\small x(\tau)$ as viewed through the sliding window $\small h(t-\tau)$. Using the real and the imaginary components, we create a signature as follows.
\begin{equation}
\label{eq:eq02}
\footnotesize \mathcal{S} = \left[f\left\lbrace\mathbb{R}(X)\right\rbrace~~~~f\left\lbrace\mathbb{I}(X)\right\rbrace~~~~\mathbb{P}\right]
\end{equation}
where $\small f\left\lbrace x\right\rbrace$ is a transformation of $\small x$; $\small \mathbb{R}(X)$ and $\small \mathbb{I}(X)$ are respectively, the real and the imaginary components of $\small X$ (Equation~\ref{eq:eq01}); and $\mathbb{P}$ is the padded zero component, which is introduced for future extension of the approach to incorporate additional parameters. The signature generated using the short-time Fourier transform components are shown in the fourth subplot of Figure~\ref{fig:ecg_motvation}, where the unused padded components $\small \mathbb{P}$ (bottom right corner in the figure) are indicated with circles. Each signature is a $82\times82$ matrix with only few elements having values above a certain threshold (sparsity $< 1$\%). These elements are shown as different color dots for each ECG sparse distributed signature. The position of these high value elements in the matrix distinguishes one signature from another uniquely, resulting in high classification accuracy and high noise tolerance \cite{donoho2006stable} (see Section~\ref{sec:results}).

\begin{figure}[t]
	\centering
	\centerline{\includegraphics[width=0.95\columnwidth]{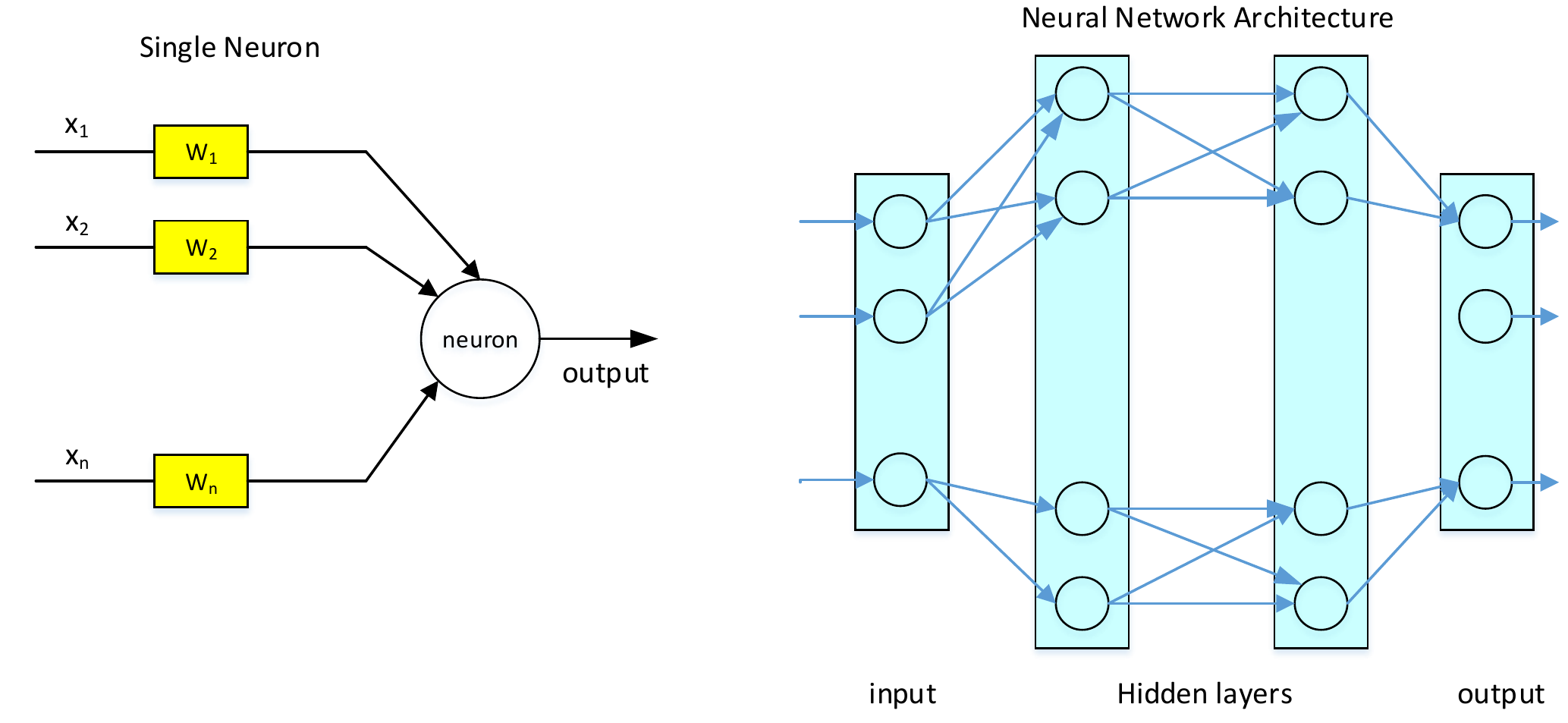}}
	\caption{Single neuron model and neural network architecture.}
	\label{fig:neural_network}
\end{figure}

\subsection{MLP Classifier}
An artificial neural network is a computing paradigm inspired by the biological computation of the brain. It comprises of processing elements (neurons) working in unison to solve a specific problem. A simple artificial neuron is shown in Figure~\ref{fig:neural_network}. It computes the weighted sum of its inputs (from external world or from other neurons). If this sum is greater than a threshold, the neuron is said to \emph{fire} (i.e., produces an output). This output is then communicated to external world or connected to the input of another neuron. A typical feed-forward neural network is shown on the right side of Figure~\ref{fig:neural_network}. This consists of an input layer, one or more hidden layers, and an output layer. The number of neurons in the input layer is equal to the dimensions (also called \emph{features}) of the input applied to this neural network. We use ECG sparse distributed signatures $\small\mathcal{S}^{82\times82}$ as input to the neural network and therefore, the number of input layer neurons is $\small 6724~(= 82\times82)$. The number of hidden layers and neurons per hidden layer are configurable and determines the training time and accuracy of classification. Finally, the number of neurons in the output layer is same as the number of heartbeat classes. It is to be noted that conceptually, the deep layers of a neural network represents hierarchical and non-linear combination of features detected from the input. Thus, instead of hand coding essential features, neural network autonomously selects (and combines) features resulting in high classification accuracy. This justifies our choice of neural networks for heartbeat classification.

Neural networks have two modes of operation -- training and testing, facilitating supervised learning. In the training mode, a neural network works with a set of inputs (ECG sparse distributed signatures) and their known outputs (QRS labels). In this mode, weights (also called \emph{synaptic weights}) of the network are adjusted to minimize the error between the expected and actual output. In the testing mode, the network output is evaluated for an unknown input. Back propagation algorithm \cite{williams1986learning} is one of the most popular approaches for supervised learning in a neural network. 

\begin{figure}[t]
	\centering
	\centerline{\includegraphics[width=0.99\columnwidth]{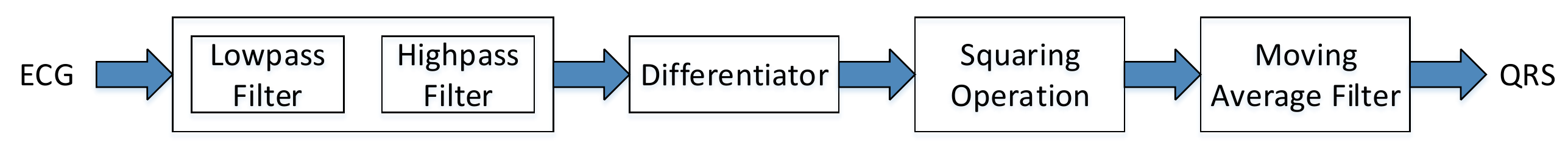}}
	\caption{QRS detection using Pan \& Tompkin's Algorithm.}
	\label{fig:qrs_detection}
\end{figure}

\subsection{QRS Detection}
A real-time implementation of Pan \& Tompkin's algorithm is proposed in \cite{pan1985real}. This is shown as a flow chart in Figure~\ref{fig:qrs_detection}. It detects QRS complexes based on the slope, amplitude and width of ECG lead II signals. The digitized signal from an acquisition system~\cite{konijnenburg201628} is first processed using a band-pass filter, which is a cascaded lowpass and highpass filter. The purpose of this filter is to reduce noise in ECG signals (from muscle contraction-expansion and T-wave interference) by matching the spectrum of average QRS complexes. Output of the bandpass filter is processed using a five point differentiator to extract information on the QRS complex and eliminate baseline drift effects. The next operation is squaring, which is a nonlinear transformation of the signal and serves two purposes -- (1) makes all data positive and (2) emphasize (and attenuate) high (and low) frequency components of the signal. The final transformation is the moving average filtering, which smooths the signal and performs moving window integration of 150ms.

\section{Results and Discussions}
\label{sec:results}
We validated our approach using the MIT-BIH Arrhythmia Database \cite{moody2001impact} (part of Physionet Database). Offline phase of our approach is implemented using python: short-time Fourier transform using \texttt{specgram} of \texttt{python-matplotlib}; and neural network training using \texttt{lasagne} of \texttt{python-theano}\cite{bergstra2011theano}. There are 23 heartbeat classes (provided as annotations in \cite{moody2001impact}). So the output layer of the neural network consists of 23 neurons. The training set is composed of 6500 labeled QRS complexes selected randomly from 120K QRS complexes for 48 patients in the database, with representatives from every classes. The test set is composed of all 120K QRS complexes (except the ones used for the training set). These test QRS complexes are represented from all 48 patients and 23 classes. Each ECG QRS complex is a sparse distributed signature of 82x82 matrix and therefore, the number of neurons in the input layer is 6724. The design space exploration determines the number of hidden layers, neurons per hidden layer and synaptic weights on nvidia GeForce GTX 970 GPU @ 1.05GHz. The onlne phase is performed on Hashwell CPU.

\begin{figure}[t]
	\centering
	\centerline{\includegraphics[width=0.80\columnwidth]{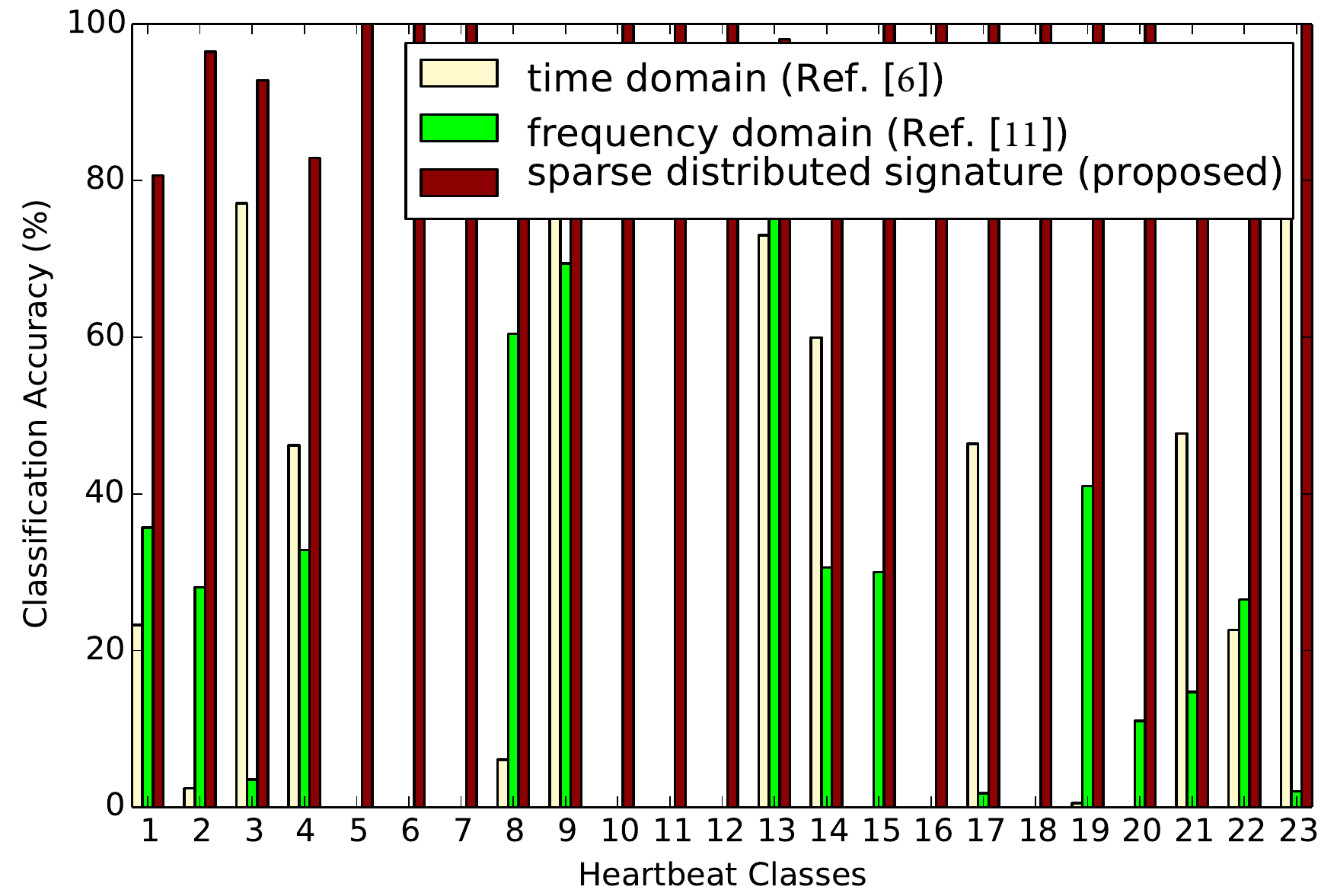}}
	\caption{Prediction results for 23 heartbeat classes.}
	\label{fig:accuracy_type}
\end{figure}

\subsection{Classification Accuracy}
Figure~\ref{fig:accuracy_type} reports performance of our approach for 23 heartbeat classes. For comparison, classification results using ECG time samples \cite{queiroz2015automatic} and ECG frequency components \cite{martis2014computer} are also reported in the plot. As can be seen from this figure, using time samples has higher classification accuracy than using frequency components for some classes such as 3, 14, 17, 21 \& 23. For these heartbeat classes, amplitude of ECG signals is sufficient for classification; frequency components convey little or no information as discussed in \cite{park2014pchd}. For other classes such as 2, 8, 15, 19 \& 20, frequency domain representation carries more information needed for classification. Compared to these approaches, our approach not only captures both time and frequency representations, but also time variations of the frequency components in the sparse representation, leading to higher classification accuracy for all classes. The accuracy obtained using our approach is 80.7\% to 100\% (average 95.7\%), $\approx$ 15\% -- 70\% (average 46\%) improvement over \cite{martis2014computer} and 2.5\% -- 54.4\% (average 22\%) improvement over \cite{queiroz2015automatic}.

\begin{table}[t]
	\renewcommand{\arraystretch}{1.2}
	\setlength{\tabcolsep}{1.2pt}
	\caption{Classification accuracy with different training classes.}
	\label{tab:class_accuracy}
	\centering
	{\fontsize{6}{8}\selectfont
		\begin{tabular}{|c|c|c|}
			\hline
			\textbf{Heartbeat Class} & \textbf{Training with 23 classes} & \textbf{Training with 5 classes}\\
			\hline
			normal & 91.7\% & 95.1\% \\ 
			superventricular & 95.3\% & 95.7\% \\
			ventricular & 98.9\% & 99.5\%\\
			fusion & 89.8\% & 93.4\%\\
			unclassified & 86.8\% & 91.2\%\\
			\hline
		\end{tabular}}
	\end{table}
	
	Table~\ref{tab:class_accuracy} reports results for the five required classes for two scenarios -- (1) when the classifier is trained with all 23 classes (column 2) and (2) when the classifier is trained with only 5 classes (column 3). As can be seen from this table, training with 5 classes improves accuracy by 0.4\% -- 5.1\% (average 2.8\%) as compared to training with all 23 classes. These results reconfirm that using ECG sparse distributed signatures result in high classification accuracy, which can be improved further by training with only the heartbeat classes required for detecting a specific heart abnormality. Thus the approach proposed in this work is generic and can be extended using disease-specific learning.
	
	\subsection{Choice of ECG Preprocessing}
	Table~\ref{tab:spectrogram_justify} reports results justifying our choice of ECG sparse distributed signatures for classification. We report 4 numbers in the table -- average accuracy for 48 patients in the database (column 2); average accuracy for 23 heartbeat classes (column 3); average false positives (column 4); and average false negatives (column 5). In this context, false positive refers to a scenario where a heartbeat is actually normal (class 1) but the classifier classifies it as abnormal, belonging to another class (classes 2 -- 23). A false negative is a scenario in which a heartbeat belongs to a specific class and the classifier classifies this to be of different class. It is to be noted that a false positive implies false alarm, while a false negative refers to miss detection. Therefore, lower are the false positives (and negatives), better is the quality.
	
	\begin{table}[t]
		\renewcommand{\arraystretch}{1.2}
		\setlength{\tabcolsep}{1.2pt}
		\caption{Justifying the choice of ECG Sparse Distributed Signatures.}
		\label{tab:spectrogram_justify}
		\centering
		{\fontsize{6}{8}\selectfont
			\begin{tabular}{|c|c|c|c|c|}
				\hline
				\textbf{Techniques} & \textbf{Patient (Avg.)} & \textbf{Type (Avg.)}  & \textbf{False +ve} & \textbf{False -ve}\\
				\hline
				Wavelet Transform\cite{shufni2015ecg} & 58.4\% & 57.4\% & 23.8\% & 43.5\%\\
				Short-time Fourier Transform & 77.1\% & 65.5\% & 18.5\% & 35.3\%\\
				Numenta SDR \cite{purdy2016encoding} & 80.3\% & 85.1\% & 21.5\% & 15.6\%\\
				\hline
				Our Sparse Signature & 83.7\% & 95.7\% & 19.3\% & 3.7\%\\
				\hline
			\end{tabular}}
		\end{table}
		
		We  compare our ECG sparse distributed signatures with (1) wavelet transform based representation \cite{shufni2015ecg}, (2) short-time Fourier transform based representation and (3) Numenta's sparse representation technique for ECG time sample \cite{purdy2016encoding}. The classifier in all these scenarios are the same (i.e., multi-layer perceptron), only the input to the classifier are different according to representations used for ECG. As seen from the table, ECG sparse distributed signature-based representation has higher accuracy than the other three representations, for all heartbeat classes and also for all patients in the database. Furthermore, ECG sparse distributed signature-based representation has lower false positives than \cite{shufni2015ecg}, but higher than using short time Fourier transform representation by only 5\%. However, our approach has a much lower false negatives (3.7\%), a reduction of 89\% and 91\% compared to the wavelet and short-time Fourier transform representations, respectively. These improvements are due to the fact that we combine the magnitude and phase of time-frequency joint distribution into sparse distributed signatures, which uniquely distinguishes QRS complexes, resulting in higher classification accuracy.
		
		\begin{figure}[t]
			\centering
			\centerline{\includegraphics[width=0.9\columnwidth]{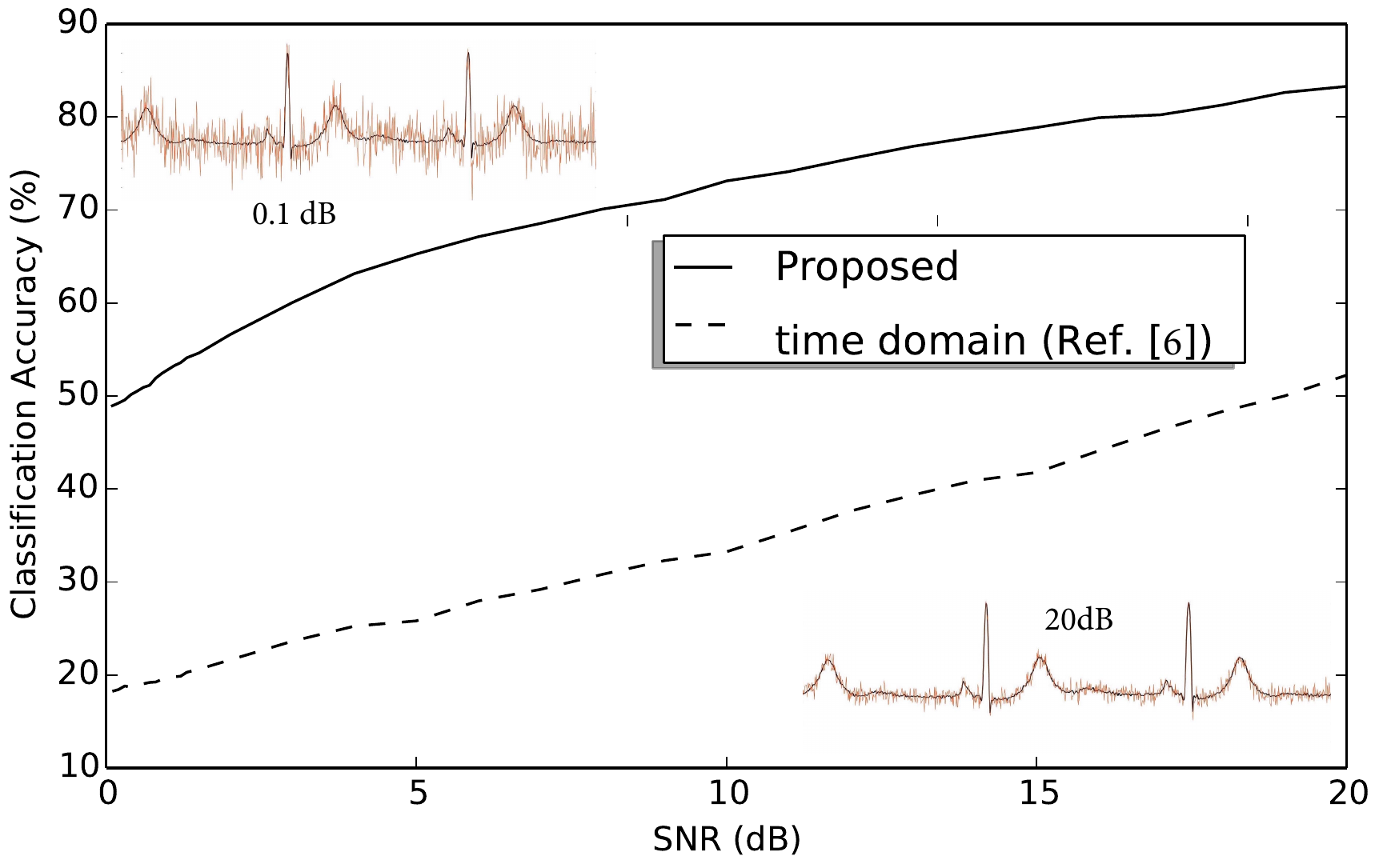}}
			\caption{Noise tolerance of the proposed approach.}
			\label{fig:noise_tolerance}
		\end{figure}
		
		\vspace{-5pt}
		
		\subsection{Additive White Noise Tolerance}
		To determine the noise tolerance of our approach, we injected random noise to the ECG signal and used our approach to classify heartbeats of this noisy ECG signal. We used Additive White Gaussian Noise (AWGN) as the basic noise model to mimic the effect of random processes associated with ECG data acquisition. Under the assumption of AWGN, let $\small X =\left\lbrace x_1,\cdots,x_L\right\rbrace$, be $\small L$ samples of the input vector $\small X$. The relative power of the noise is described in terms of signal-to-noise ratio (in dB) per sample. Let this be $\small SNR$. Power of the input ECG signal (without noise) is
		\begin{equation}
		\footnotesize E_s = \frac{1}{L}\sum_{i=1}^{L} |x_i|^2
		\end{equation}
		The noise vector is given by
		\begin{equation}
		\footnotesize N_s = \sqrt{\frac{E_s}{SNR_{lin}}}\cdot randn(1,L)
		\end{equation}
		where, $\small SNR_{lin} = 10^{SNR/10}$ is the SNR value converted to linear scale, and $\small randn(1,L)$ is a function that generates $\small L$ random numbers. The noisy ECG signal is $\small X_n = X + N_s$.
		
		Figure~\ref{fig:noise_tolerance} plots the classification accuracy of our approach, as the SNR per sample is increased from 0 to 20 dB, for patient id 106 in the database. To compare the noise tolerance, we have also plotted classification results using ECG time samples \cite{queiroz2015automatic}. As shown, the accuracy of our approach and that of \cite{queiroz2015automatic} increases with an increase in the SNR as signal strength increasingly dominates over noise, resulting in higher classification accuracy. A similar trend is observed for all other patients in the database. Additionally, our approach has higher classification accuracy (60\% -- 170\%, average 133\% for the SNR range of 0 to 20 dB) compared to \cite{queiroz2015automatic}. This is because sparse distributed signatures can tolerate high noise than other representations.
		
		Finally, motion artifacts represent an important class of noise for ambulatory subjects. Our initial experiments with motion compensated ECG from \cite{behravan2015rate} provides a comparable accuracy with recently reported wavelet-ICA approach \cite{abbaspour2015ecg}. 
		
		\begin{table}[t]
			\renewcommand{\arraystretch}{1.4}
			\setlength{\tabcolsep}{1.2pt}
			\caption{Offline and online execution times.}
			\label{tab:real_time}
			\centering
			{\fontsize{6}{8}\selectfont
				\begin{tabular}{|c|c|c|c|c|}
					\hline
					\multirow{2}{*}{\textbf{Neurons}} & \textbf{Offline} & \multicolumn{3}{|c|}{\textbf{Online (real-time) at 160 MHz}} \\ 
					\cline{2-5}
					& \textbf{Training} & \textbf{QRS Detect}  & \textbf{Sparse Distributed Signature} & \textbf{Classification} \\
					\hline
					2000 & 620.4 s & 10 $\mu$s& 10.4 ms & 1.20 ms\\
					4000 & 907.9 s & 10 $\mu$s& 10.4 ms & 1.70 ms\\
					6000 & 1277.9 s & 10 $\mu$s& 10.4 ms & 2.42 ms\\
					8000 & 1693.3 s & 10 $\mu$s& 10.4 ms & 3.06 ms\\
					10000 & 2245.4 s& 10 $\mu$s& 10.4 ms & 3.97 ms\\
					\hline
				\end{tabular}}
			\end{table}
			
			\subsection{Real-time Performance}
			Table~\ref{tab:real_time} reports execution time for the offline training phase and three components of the online phase -- the QRS detection, ECG sparse distributed signature conversion and heartbeat classification. These are reported in columns 2, 3, 4 and 5, respectively for different number of neurons. It is to be noted that the offline phase is executed on nvidia GPU and does not contribute to real-time performance. As seen from column 2, the offline training time increases exponentially with increase in the number of neurons. This is due to an exponential increase in the number of synapses, leading to an increase in the number of error computations in the back propagation algorithm. 
			At run-time, the QRS detection algorithm takes on average 10 $\mu$s to execute the steps listed in Figure~\ref{fig:qrs_detection} and are reported in column 3. The execution time to convert a QRS peak to ECG sparse distributed signature is reported in column 4 and is on average 10.4 ms.  Finally, the execution time for the heartbeat classification step is reported in column 5. As shown in this column, heartbeat classification using a trained neural network increases with  the number of neurons. However, the small execution time validates the real-time property of the proposed approach.
			
			%
			
			\section{Conclusion}
			\label{sec:conclusion}
			We proposed a real-time approach for classifying heartbeats to detection abnormal heart conditions such as arrhythmia. The approach is based on multi-layer perceptron classifier that uses ECG sparse distributed signatures to classify incoming ECG heartbeats into one of 23 different classes. Experiments with ECG signals from MIT-BIH arrhythmia database show that the proposed approach has a classification accuracy of an average 95.7\% (22\% better than state-of-the-art). Additionally, the proposed ECG sparse distributed signatures have higher noise tolerance and reduce false negatives to 3.7\%, an 89\% improvement over existing ECG data representation techniques.

\begin{acks}
	This work is supported in parts by EU-H2020 grant NeuRAM3 Cube (NEUral computing aRchitectures in Advanced Monolithic 3D-VLSI nano-technologies) and ITEA3 proposal PARTNER (Patient-care Advancement with Responsive Technologies aNd Engagement togetheR). 
\end{acks}

\bibliographystyle{IEEEtran}
\balance
\bibliography{ecg}

\end{document}